\documentclass[12pt,preprint]{aastex}

\newcommand\simlt{\lower.5ex\hbox{$\; \buildrel < \over \sim \;$}}
\newcommand\simgt{\lower.5ex\hbox{$\; \buildrel > \over \sim \;$}}

\begin{document}
\title{Ultra-relativistic, neutrino driven flows in GRBs: A double transonic flow solution in Schwarzschild spacetime}
\author{Amir Levinson$^{1}$ and Noemie Globus$^{1}$}
\altaffiltext{1}{School of Physics \& Astronomy, Tel Aviv University, Tel Aviv 69978, Israel}

\begin{abstract}
The structure of a hydrodynamic, double transonic flow driven by neutrino annihilation in the polar region of a Schwarzschild black hole is computed for
different energy deposition profiles.  The requirement that both, the inflow into the black hole and the outflow to infinity pass smoothly 
through their sonic points fixes the stagnation radius and stagnation pressure.   The asymptotic power of the outflow is shown to be the integral of the 
energy deposition rate above the stagnation radius.   The outflow production efficiency depends on the energy deposition profile, and is generally higher for shallower profiles.   Using recent calculations of the neutrino annihilation rate, we estimate that over 50 percents of the total energy deposited above the horizon can emerge in the form of a relativistic outflow at infinity.   The continuous creation of plasma during the expansion of the outflow leads to generation of a large specific entropy.  This has important implications for the prompt photospheric emission in GRBs.

\end{abstract}

\section{Introduction}

The relativistic outflows producing the prompt and afterglow emissions  in GRBs are commonly thought to be powered by hyperaccreting black holes. 
The high Lorentz factors inferred from energy considerations, compactness arguments and afterglow models, $\Gamma\sim10^2 - 10 ^3$,  require
extremely low baryon load at the outflow injection point, which seems difficult to achieve in disk 
outflows (Levinson 2006; Barzilay \& Levinson 2008; Metzger et al. 2008). 
A natural way to avoid baryonic contamination is to accelerate the outflow in the region above the horizon of the black hole (Levinson \& Eichler, 1993).
Two popular jet production mechanisms, that are widely discussed in the literature, are magnetic extraction of the rotational energy of a Kerr black hole (Blandford \& Znajek 1977),  
and pair creation on horizon threading field lines by annihilation of neutrinos that emanate from the accretion disk surrounding the black hole (Eichler et al. 1989, Popham et al. 1999; Asano \& Fukuyama 2001; Birkl et al. 2007; Zalamea and Beloborodov 2011; hereafter ZB11).   The former mechanism requires nearly maximal rotation of the black hole and sufficiently high magnetization. The latter mechanism can operate also in a Schwarzschild spacetime, although extremely high accretion rates ($\simgt 3 M_\odot$ s$^{-1}$) are required in this case  to account for the characteristic GRB luminosities (ZB11).  It is worth noting that the presence of a rapidly spinning Kerr black hole may not be sufficient to guarantee effective production of a  Blandford-Znajek (hereafter BZ) outflow; overloading of magnetic field lines by relativistic pairs may lead to a complete shutdown of this mechanism, in which case one has to rely on the neutrino-driven outflow to power a GRB.  However, 
the accretion rate required to power a typical GRB by the latter mechanism can be considerably reduced when the specific angular 
momentum of the black hole, $a$, is high enough. 
The reason is that the neutrino luminosity emitted by the disk and, hence, the rate at which energy is deposited above the horizon by $\nu\bar{\nu}$ annihilation, 
increases sharply with increasing $a$ (ZB11). 

A question of interest is what fraction of the total energy deposited above the horizon via $\nu\bar{\nu}$ annihilation emerges at infinity.  To address this question, we constructed a model for the double-transonic flow established in the polar region, that incorporates energy injection by the external neutrino source in a self-consistent manner.   

\section{\label{sec:model}Model description}

\begin{figure}[ht]
\centering
\includegraphics[width=12cm]{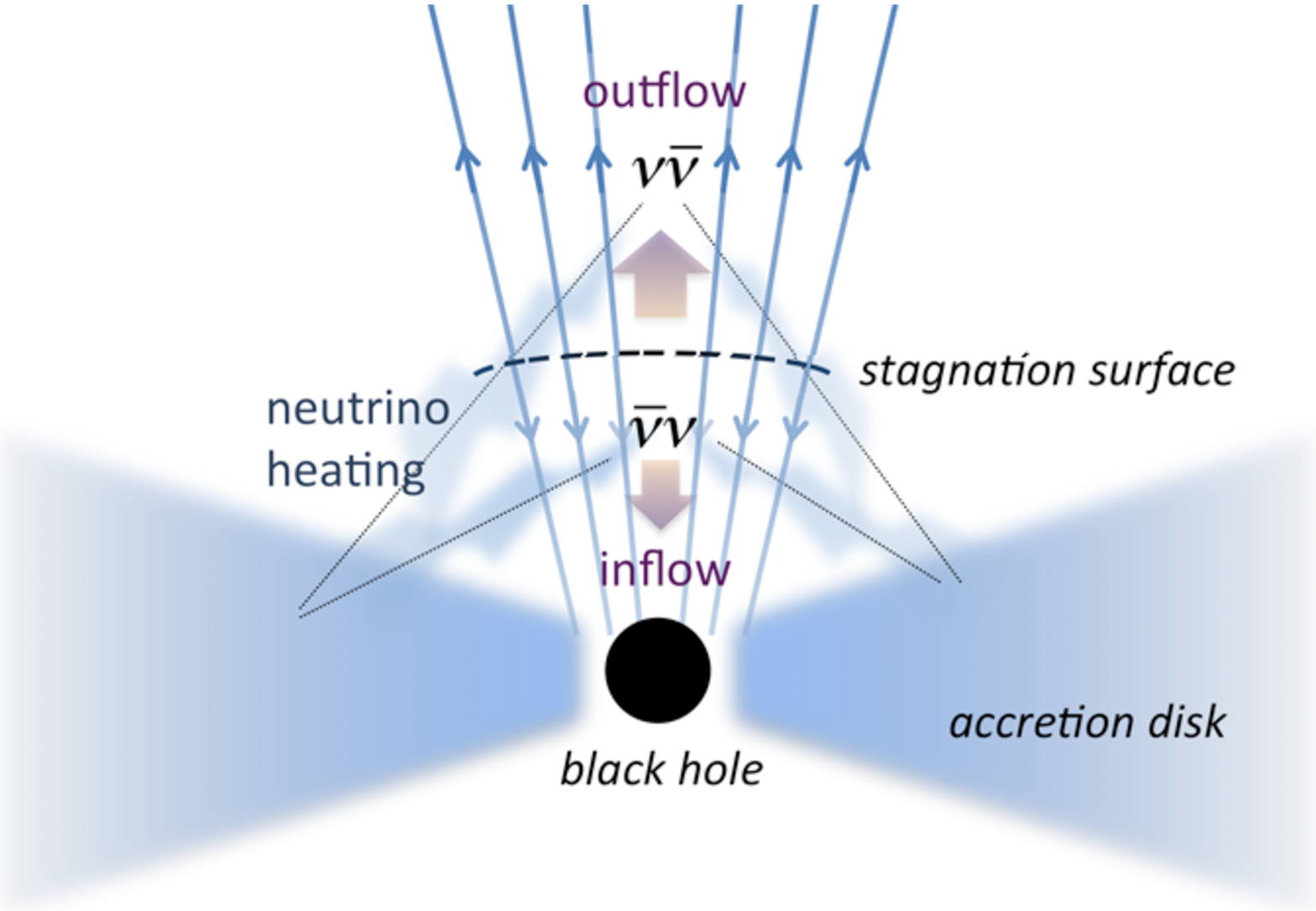}
\caption{\label{fig:schem} Schematic illustration of the double-transonic flow model.}
 \end{figure}

Neutrino annihilation in the baryon free region above the horizon generates a relativistically hot fluid consisting of a mixture of e$^\pm$ pairs and radiation in equilibrium.  Part of this fluid is accelerated outwards by pressure gradient forces, and the other part falls into the black hole, owing to the strong
gravitational force exerted on it.  In what follows, we consider the structure of the double transonic flow thereby produced. Schematic illustration of the model
is shown in figure \ref{fig:schem}.
We restrict our analysis to the Schwarzschild spacetime, and suppose that the flow is stationary, radial, unmagnetized and non-rotating.  For clarity we consider baryon free plasma, although our analysis applies also when baryons are present, provided the enthalpy per baryon is sufficiently large ($w/n_b>>1$).  
The idea is to find a self-consistent solution that starts from a stagnation point, where the flow velocity vanishes, and crosses two sonic points; an inner one located in the inflow section below the stagnation point,  and an outer one located in the outflow section above the stagnation point.   As shown below, 
for a given choice of energy deposition profile the stagnation radius and stagnation pressure are uniquely determined by the requirement that the solution passes smoothly through both sonic points.

\subsection{\label{sec:injection} Energy deposition rate}
The annihilation of neutrinos emitted from a hyper-accretion disk was calculated in a number of recent works, under different simplifying 
assumptions (e.g., Popham et al. 1999; Asano \& Fukuyama 2001; Birkl et al. 2007; ZB11).
The most detailed analysis is presented in ZB11, who adopted the relativistic disk model of Chen and Beloborodov (2007) as the neutrino source, and  
employed a geodesic-tracing method in Kerr spacetime to evaluate the local energy-momentum deposition rates by the reaction $\nu\bar{\nu}\rightarrow$ e$^+$e$^-$ in the vicinity of the black hole. 

The local energy-momentum deposition rates, $Q_{\nu\bar{\nu}}^\alpha$, were computed in ZB11 in the frame of a zero angular momentum observer (ZAMO).  
Quite generally, these rates are functions of  $r$ and $\theta$, when expressed in Boyer-Lindquist  coordinates, $x^u=(t,r,\theta,\phi)$.    Since we restrict 
our analysis to a conical flow, we shall adopt  a local energy deposition rate of the form:
\begin{equation}
Q^t_{\nu\bar{\nu}}(r)=\dot{Q}_0f(r/r_H),
\label{Qdot}
\end{equation}
where $Q^t_{\nu\bar{\nu}}(r)$ represents the angle-averaged rate at radius $r$, $r_H$ is the horizon scale, and $f(1)=1$.  From figures 2 and 3 in  ZB11 we estimate $f(x)\simeq x^{-4.5}$ for $a=0.95$ and $f(x)\simeq x^{-3.5}$ for $a=0$, in the range $r_H\le r\le 30m$ delineated in the figures, where $m$ is the black hole mass in geometrical units and $a$ its specific angular momentum.   The profile should steepen as the radius increases,  approaching  $f(x)\propto x^{-8}$ at radii much larger than the size $R_\nu$ of the neutrino source (e.g., Goodman et al. 1987, Qian \& Woosley 1996).   For the disk model outlined in Chen \& Beloborodov (2007), $R_\nu$ can be identified with the ignition radius of the disk, $r_{ign}$, within which neutrino emission is efficient. 
 
The net energy deposition rate, as measured by a distant observer, is given in terms of the ZAMO rates $Q_{\nu\bar{\nu}}^\alpha$ by (ZB11)
\begin{equation}
\dot{E}_{\nu\bar{\nu}}=\int_{r\ge r_H}{\left(\frac{Q^t_{\nu\bar{\nu}}}{\sqrt{-g^{tt}}}-\frac{g_{t\phi}}{\sqrt{g_{\phi\phi}}}Q_{\nu\bar{\nu}}^\phi\right)\sqrt{-g}dr d\theta d\phi},
\label{Edot}
\end{equation}
here $g_{\mu\nu}$ are the metric components of the Kerr spacetime in Boyer-Lindquist coordinates.   The results exhibited in figure (4) of ZB11 (see also their Equation (22) for a fitting formula) indicate that $\dot{E}_{\nu\bar{\nu}}$ depends sensitively on the accretion rate $\dot{m}$ (given in units of $M_\odot /s$) and the spin of the black hole $a$, with  $\dot{E}_{\nu\bar{\nu}}\simeq5\times10^{49}\dot{m}^{9/4}$ erg s$^{-1}$at $a=0$ and $\dot{E}_{\nu\bar{\nu}}\simeq 10^{52}\dot{m}^{9/4}$ erg s$^{-1}$ at $a=0.95$, for a  black hole mass $M_{BH}=3M_\odot$, and accretion rates in the range $\dot{M}_{ign}<\dot{M}<\dot{M}_{trap}$.  At $\dot{M}<\dot{M}_{ign}$ neutrino emission is 
severely suppressed by virtue of the low disk temperature.  At $\dot{M}_{trap}<\dot{M}$ the neutrinos are trapped in the disk and advected into the black hole, whereby the neutrino luminosity saturates.

\subsection{Flow equations}

The stress-energy tensor of a purely hydrodynamic flow takes the form
\begin{equation}
T^{\alpha\beta} = wu^{\alpha}u^{\beta} + pg^{\alpha\beta},
\label{T_M}
\end{equation}
here $u^{\alpha}$ is the four-velocity measured in units of c, $w$ is the specific enthalpy, $p$ the pressure, and $g_{\mu\nu}={\rm diag}(-\alpha^2,\alpha^{-2}, r^2, r^2\sin^2\theta)$, $\alpha^2=1-2m/r$, is the metric tensor of the  Schwarzschild spacetime. 

The dynamics of the flow is governed by the energy-momentum equations:
\begin{equation}
\frac{1}{\sqrt{-g}}(\sqrt{-g}T^{\alpha\beta})_{,\alpha}+
\Gamma^{\beta}_{\ \mu\nu}T^{\mu\nu} = q^\beta, \label{Ttot=q}
\end{equation}
where $q^{\beta}$ denotes the source terms associated with energy-momentum transfer by the external agent, and $\Gamma^\beta_{\mu\nu}$ denotes the affine connection.
By contracting $g_{\beta\gamma}$ with Equation (\ref{Ttot=q}), using the relation $(\sqrt{-g}g^{\alpha\beta})_{,\alpha}+\sqrt{-g}
\Gamma^\beta_{\mu\nu}g^{\mu\nu}=0$, taking the $t$ component and noting that $\Gamma_{\mu t \nu}u^\mu u^\nu=0$ for a stationary flow, one obtains
\begin{equation}
\frac{1}{\sqrt{-g}}(\sqrt{-g}w u^{\alpha} u_t)_{,\alpha}=q_t.
\label{energy}
\end{equation}
Likewise, contracting $u_\beta$ with Equation (\ref{Ttot=q}), using the identity 
$\Gamma^{\beta}_{\ \mu\nu}u_\beta u^\mu u^\nu=-u_\beta u^\nu(u^\beta)_{,\nu}$, yields
\begin{equation}
\frac{1}{\sqrt{-g}}(\sqrt{-g}w u^{\alpha} )_{,\alpha}-u^\alpha p_{,\alpha}=-u_\nu q^\nu.
\label{entropy}
\end{equation}
Combining Equation (\ref{entropy}) with the thermodynamic identity $d(wV)-Vdp=kT d(Vs)$, where $V$ is the volume of a fluid element, $T$ its temperature and and $s$ its specific entropy, yields the change in $s$:
\begin{equation}
\frac{kT}{\sqrt{-g}}(\sqrt{-g}s u^{\alpha} )_{,\alpha}=-u_\nu q^\nu.
\label{entropy2}
\end{equation}

Henceforth, we consider a conical flow, for which $u^\mu=(u^t,u^r,0,0)$, $u^\mu\partial_\mu=u^r\partial_r$, and denote by $u_p=u^r/\alpha$ the poloidal velocity, 
and by $\gamma=(1+u_p^2)^{1/2}$ and $v=u_p/\gamma$ the corresponding Lorentz factor and 3-velocity.  The convention here is
that $u_p<0$ on inflow lines and $u_p>0$ on outflow lines.   Since the created pairs are relativistic, we adopt the equation of state $w=4p$.  Furthermore, we neglect momentum transfer to the flow by the created pairs, so that $q^\beta=(q^t,0,0,0)$.  The source term $q^t$ is measured in the frame of a distant observer, and can be expressed in terms of the local energy deposition rate adopted in Equation (\ref{Qdot}) as $cq^t=\sqrt{-g^{tt}}Q^t_{\nu\bar{\nu}}=\dot{Q}_0f(x)/\alpha$. 
Under the above simplifications, Equations (\ref{energy}) and (\ref{entropy}) reduce to
\begin{equation}
(3v^2-1)\partial_x\ln u_p=\frac{(3-4\gamma^2)}{\gamma u_p 4\tilde{p}}\frac{f(x)}{(1-x^{-1})^{1/2}}+\frac{2}{x}-\frac{1}{x(x-1)},
\label{eq-motion}
\end{equation}
and 
\begin{equation}
\partial_x\ln \tilde{p}=\frac{f(x)}{\gamma u_p 4\tilde{p}(1-x^{-1})^{1/2}}-\frac{2}{x}-\frac{1}{x(x-1)}-(1+v^2)\partial_x\ln u_p,
\label{eq-pressure}
\end{equation}
in terms of the dimensionless radius $x=r/r_s$, $r_s=2m$, and the normalized pressure $\tilde{p}=p/(\dot{Q}_0t_d)$, $t_d=r_s/c$ 
being the dynamical time of the black hole. Equation (\ref{eq-motion}) has critical points at $v=\pm c_s$, 
where $c_s=1/\sqrt{3}$ is the sound speed.

At the stagnation point $x=x_{st}$, where $u_p=0$, the above equations yield 
\begin{eqnarray}
\partial_x u_p=\frac{f(x_{st})}{4 \tilde{p}_{st}(1-x_{st}^{-1})^{1/2}},\\
\partial_x\ln(\tilde{p})=-\frac{2}{x_{st}(x_{st}-1)},
\label{stagnation}
\end{eqnarray}
where $\tilde{p}_{st}=\tilde{p}(x_{st})$ is the stagnation pressure.  Thus, 
for a given choice of $f(x)$ the solution is fully determined once $x_{st}$ and $p_{st}$ are known. 

The regularity conditions at the sonic points, obtained from Equation (\ref{eq-motion}), read
\begin{equation}
\sqrt{3}x^2_{c1}(1-x^{-1}_{c1})^{1/2}f(x_{c1})=-2(2x_{c1}-3)\tilde{p}_{c1},\label{x_c1}
\end{equation}
and
\begin{equation}
\sqrt{3}x^2_{c2}(1-x^{-1}_{c2})^{1/2}f(x_{c2})=2(2x_{c2}-3)\tilde{p}_{c2},\label{x_c2}
\end{equation}
denoting the sonic point of the inflow (outflow) by $x_{c1} (x_{c2})$, and
noting that $u_p= -1/\sqrt{2}$ at $x_{c1}$, and  $u_p= 1/\sqrt{2}$ at $x_{c2}$.  Evidently, $x_{c1}<3/2$ and $x_{c2}>3/2$.  
The existence of two sonic points is a consequence of energy injection. When $f(x)=0$ the system has only one critical point at $x_c=3/2$.

\subsection{Asymptotic power and entropy of the outflow}
Well above the sonic radius, at $x>>x_{c2}$, the power of the outflow is given by 
$L_{j\infty}=2\pi c\int_0^{\theta_0}{T^{0r}r^2\sin\theta d\theta}=2\pi(1-\cos\theta_0)w\gamma u_p$, here $\theta_0$ is the opening angle of the flow.  
Integrating Equation (\ref{energy}) from the stagnation point $x_{st}$ to infinity, using the boundary condition $u_p(x_{st})=0$ 
and noting that $q_t=g_{tt}q^t=-\alpha^2q^t$, one obtains
\begin{equation}
L_{j\infty}=2\pi(1-\cos\theta_0)\int_{r_{st}}^\infty{cq_t r^2dr}
=2\pi(1-\cos\theta_0)\dot{Q}_0r_s^3\int_{x_{st}}^\infty{(1-x^{-1})^{1/2}f(x) x^{2}dx}.
\label{L_jinf}
\end{equation}
Let us define the outflow production efficiency $\epsilon$ to be the fraction
of $\dot{E}_{\nu\bar{\nu}}$ that emerges at infinity.  From Equation (\ref{Edot}) with $g_{t\phi}=0$, $\sqrt{-g^{tt}}=\alpha^{-1}$, and Equation (\ref{L_jinf}), one finds
\begin{equation}
\epsilon\equiv \frac{L_{j\infty}}{\dot{E}_{\nu\bar{\nu}}}=\frac{\int_{x_{st}}^\infty{(1-x^{-1})^{1/2}f(x) x^{2}dx}}{{\int_{1}^\infty{(1-x^{-1})^{1/2}f(x) x^{2}dx}}}.
\label{efficiency}
\end{equation}

The rate at which entropy is carried by the flow at some radius $r$ is obtained by integrating Equation (\ref{entropy2}) from the stagnation radius $r_{st}$ to $r>r_{st}$:  
\begin{equation}
\frac{dS(r)}{dt}\equiv 2\pi(1-\cos\theta_0) r^2cu^r s=2\pi(1-\cos\theta_0)\frac{\dot{Q}_0r_s^3}{kT_0}\int_{x_{st}}^x \gamma\tilde{p}^{-1/4}f(x^\prime)x^{\prime2} dx^\prime,
\end{equation}
where $T_0$ is a fiducial temperature defined by\footnote{The pressure of a relativistically hot plasma at a temperature $T$, consisting of electrons, positrons and radiation in equilibrium, satisfies $p=11aT^4/12$, from which we obtain $T=T_0\tilde{p}^{1/4}$.} $T_0=(12\dot{Q}_0t_d/11a)^{1/4}$.  
If the outflow also carries baryons at a rate $\dot{N}_b=2\pi(1-\cos\theta_0)n_bu^rr^2$, for a baryon density $n_b$, then the dimensionless entropy per baryon can be expressed as 
\begin{equation}
\sigma(x)=\frac{1}{\dot{N}_b}\frac{dS}{dt}=\frac{m_pc^2\gamma_\infty}{kT_0}F(x),
\label{specific-s}
\end{equation}
with $\gamma_{\infty}=L_{j\infty}/(\dot{N}_bm_pc^2)$ being the maximum Lorentz factor of the outflow, and
\begin{equation}
F(x)=\frac{\int_{x_{st}}^x \gamma\tilde{p}^{-1/4}f(x^\prime)x^{\prime 2} dx^{\prime}}{\int_{x_{st}}^\infty{(1-x^{\prime -1})^{1/2}f(x^\prime) x^{\prime2}dx^\prime}}.
\label{F(x)}
\end{equation}

\section{Results}

Equations (\ref{eq-motion}) and (\ref{eq-pressure}) have been integrated numerically using energy deposition profile of the form $f(x)=x^{-b}$.   The pressure at the inner sonic point, $\tilde{p}_{c1}$, is used as a free parameter that we adjust to converge to the desired solution.   
The integration starts at the inner sonic point $x_{c1}$, where $u_p(x_{c1})=-1/\sqrt{2}$, and repeated iteratively by changing the value of $\tilde{p}_{c1}$, until a smooth transition across the outer sonic point is obtained.  The value of  $x_{c1}$ is computed, in each run, from the regularity condition (\ref{x_c1}).     The stagnation radius $x_{st}$ is then found numerically from the condition $u_p(x_{st})=0$.

Solutions are exhibited in figure \ref{fig:profile} for different values of $b$.  The horizontal dashed lines in the left panel mark the sonic velocities of the inflow ( $u_p=-1/\sqrt{2}$) and outflow ($u_p=1/\sqrt{2}$).  As seen, in all cases the sonic point of the outflow is located well above that of an adiabatic flow ($x_c=3/2$).  This is a consequence of the injection of energy over extended scales, that loads the flow and delays its acceleration.     The  values of the stagnation radii found from the integration are $x_{st}=1.57, 1.70, 1.81$ and $1.95$ for $b=5, 4, 3.5,$ and $3$, respectively.  For a comparison, note that the radius at which the escape velocity, $v_{esc}=\sqrt{r_s/r}=x^{-1/2}$, equals the sound speed, $c_s=1/\sqrt{3}$, is $x=3$, slightly above the stagnation point.  The corresponding efficiencies, computed using Equation (\ref{efficiency}), are $\epsilon=0.58, 0.73, 0.83$ for $b=5, 4, 3.5$.   For $b=3$ the energy deposition rate diverges logarithmically and needs to be cut off at some radius $x_\nu$.  We arbitrarily invoked $x_\nu=100$, whereby  $\epsilon=0.93$.  

By employing Equations (\ref{L_jinf}) and (\ref{specific-s}), we can express the entropy per baryon in the form:
\begin{equation}
\sigma(x)=1.3\times10^5\eta_b\left(\frac{\gamma_\infty}{300}\right)\left(\frac{L_{iso}}{10^{51} {\rm erg/s}}\right)^{-1/4}
\left(\frac{M_{BH}}{3 M_\odot}\right)^{1/2}F(x),
\end{equation}
where $L_{iso}=2L_{j\infty}/(1-\cos\theta_0)$ is the isotropic equivalent luminosity, $M_{BH}$ is the mass of the black hole, and 
$\eta_b=1.06, 0.84, 0.62$ for $b=3.5, 4, 5$, respectively.  A plot of $F(x)$ is shown in figure \ref{fig:entropy}.  The 
asymptotic value of $F(x)$ represents enhancement of the entropy relative to that produced in an adiabatic outflow injected from a radius $R_0\sim r_s$.

\subsection{Asymptotic behavior}
\begin{figure}[ht]
\centering
\includegraphics[width=14cm]{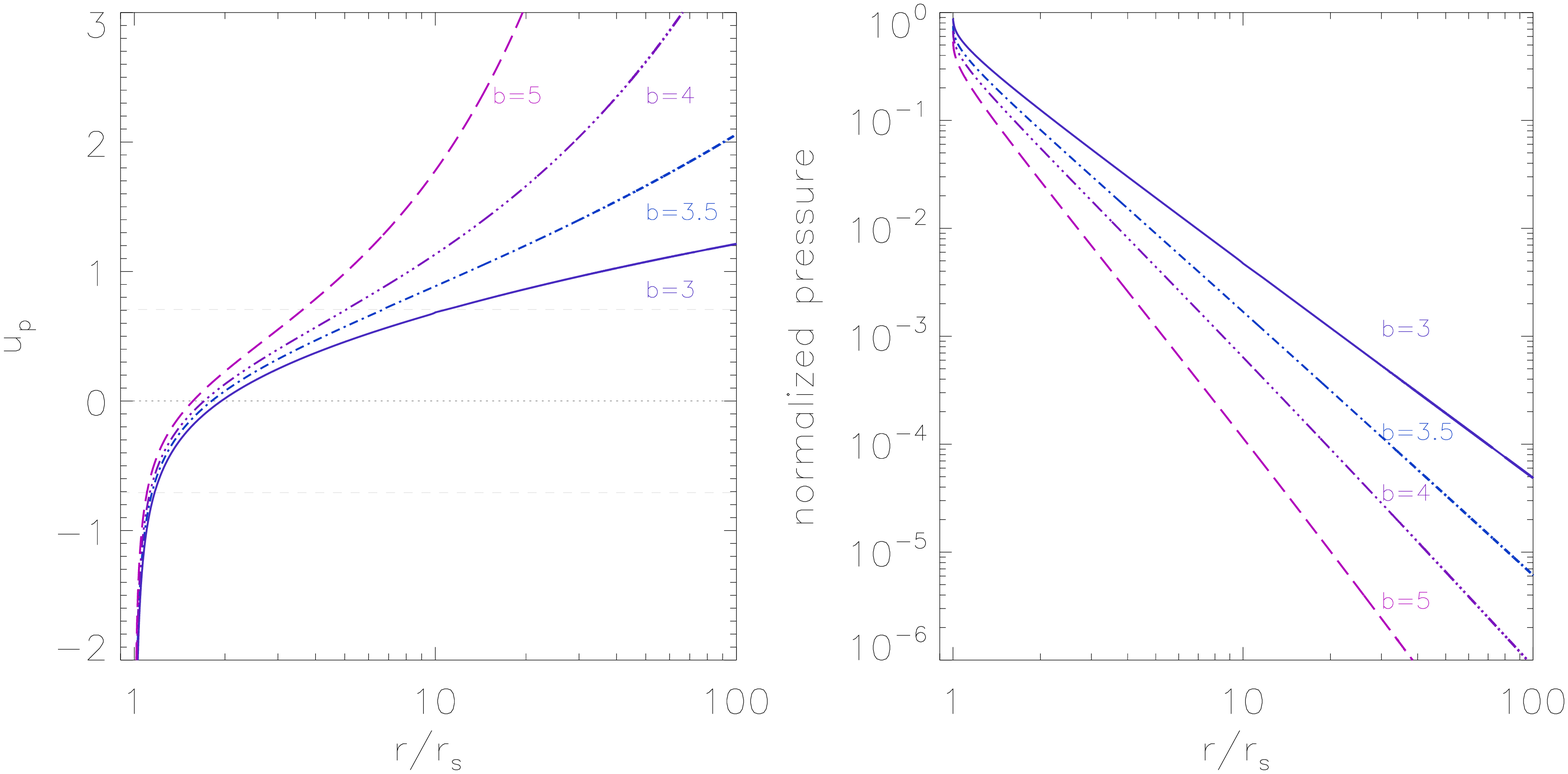}
\caption{\label{fig:profile} Velocity (left panel) and pressure (right panel) profiles for $f(x)=x^{-b}$  ($x=r/r_s$) and 
different values of $b$, as indicated.  
The region above (below) the dotted line $u_p=0$ in the left panel corresponds to the outflow (inflow).  The 
horizontal dashed lines delineate the sonic velocities of the inflow (lower line) and outflow (upper line).}
\end{figure}

As pointed out in Section \ref{sec:injection}, sufficiently far out the energy injection profile must steepen, approaching $f(x)\propto x^{-8}$ at $x>>R_\nu/2m$, where $R_\nu$ is roughly the size of the neutrino source (i.e., the disk radius within which neutrinos are emitted).  The first term on the right hand side of Equation (\ref{eq-motion}) then becomes negligibly small, and the outflow enters the adiabatic regime, whereby 
Equation (\ref{eq-motion}) can be solved analytically in the limit $v\simeq1$ and $f(x)=0$.  The solution reads: $u_p(x)\propto x^{3/2}/(x-1)^{1/2}$, and reduces to the well known result $u_p\propto x$, $\tilde{p}\propto x^{-4}$,   at $x>x_{c2}>>1$.    From Equations (\ref{eq-motion}) and (\ref{eq-pressure}) it is clear that the solution will approach this asymptotic behavior  once $b>4$, as indeed confirmed in figure \ref{fig:profile}. 

Near the horizon we obtain the analytic solution $u_p\propto -1/(x-1)^{1/2}$ for any value of $b$, in accord with the numerical solutions exhibited in figure \ref{fig:profile} , indicating that the inflow moves along radial geodesics as it approaches the horizon.  This is a consequence of the fact that near the horizon the dynamics of the inflow is dictated by the gravitational attraction of the black hole.  

\section{Conclusions}

We have computed the structure of a double transonic flow  generated  above the horizon of a Schwarzschild black hole by annihilation of neutrinos emitted from a hyperaccretion disk.  We have shown that for a given choice of the energy deposition profile, there exists a unique solution that passes smoothly through the inner and outer  sonic points.   The stagnation point was found to be located slightly below the escape radius of sonic material - in the range 1.5 - 2 Schwarzschild radii, depending on the energy deposition profile.   The asymptotic power of the outflow is given as the integral of the energy deposition rate above the stagnation radius.   The outflow production efficiency was found to be typically large, with over one half of the injected energy emerging at infinity.

\begin{figure}[ht]
\centering
\includegraphics[width=14cm]{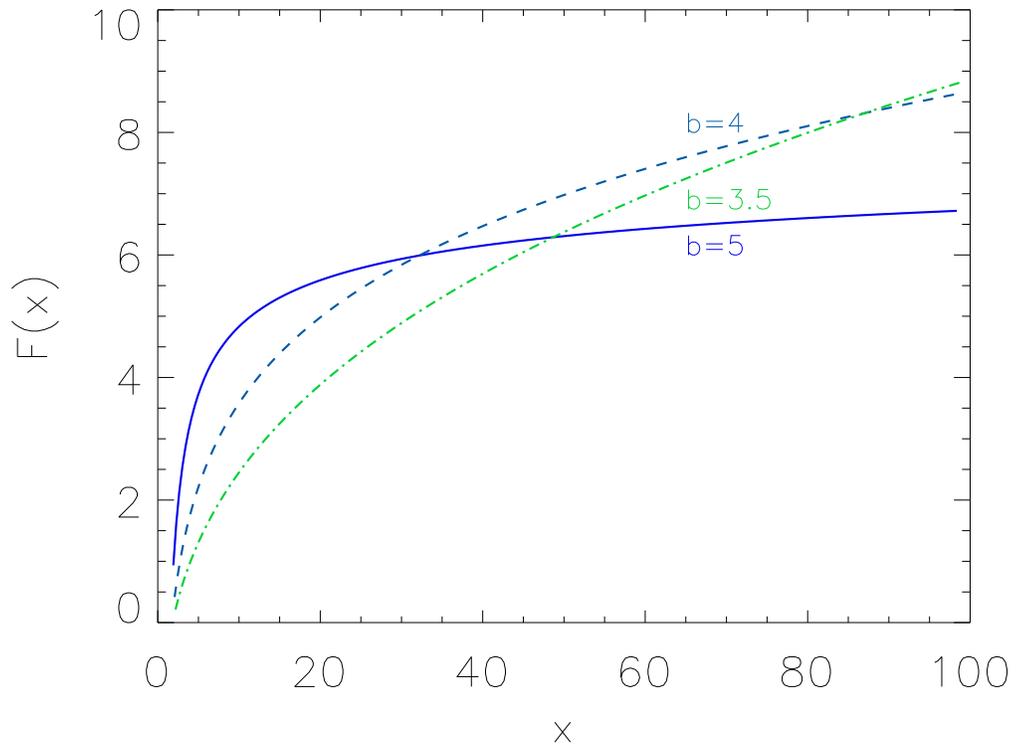}
\caption{\label{fig:entropy} A plot of the function $F(x)$ defined in Equation (\ref{F(x)}), for $f(x)=x^{-b}$ and different values of $b$, as indicated.}
 \end{figure}

The continuous injection of energy pushes out the outer sonic point and delays the acceleration of the outflow above the sonic point.  The linear acceleration phase commences once the energy deposition rate becomes small and the outflow enters the adiabatic regime.  The radius at which this happens depends on the energy deposition profile, and may exceed $100m$ at large accretion rates.   The deposition of energy over extended scales leads to generation of a relatively large specific entropy. 
As a consequence, in models wherein the prompt emission originates from the photosphere (e.g., Ryde \& Peer 2009; Lazzati et al. 2009; Peer et al. 2012; Levinson 2012; Beloborodov 2013),  the thermal peak may be located at an energy lower than previously estimated.  For instance, if the prompt emission is produced by sub-photospheric, radiation mediated shocks (Bromberg et al. 2011; Levinson 2012), than the observed temperature behind the shock is
$$
kT_{obs}\simeq50 \left(\frac{F_\infty}{10}\right)^{-1}\left(\frac{L_{iso}}{10^{51}\ {\rm erg/s}}\right)^{1/4}
\left(\frac{M_{BH}}{M_\odot}\right)^{-1/2}u_s\quad {\rm keV},
$$
where $u_s$ is the shock 4-velocity (i.e., the 4-velocity of the upstream fluid, as measured in the shock frame), and $F_\infty$ is the asymptotic value of the function $F(x)$ exhibited in figure \ref{fig:entropy}. Consequently, the peak of the spectrum produced by mildly relativistic shocks ($u_s\simgt1$) can be located at an energy $E_{peak}\sim 3kT_{obs}\sim 200$ keV,  significantly lower than those exhibited in Levinson (2012).  Additional dissipation and entropy generation may result from the interaction of the outflow with the stellar envelope, via formation of re-collimation shocks (Bromberg \& Levinson 2007).

In situations where the central black hole is rapidly rotating, the outflow can be produced by the Blandford-Znajek mechanism.   The resultant Poynting power  is expected to
exceed the net power deposited above the horizon by the reaction $\nu\bar{\nu}\rightarrow e^+ e^-$ (e.g., Kawanaka et al. 2012).   However, activation of the BZ mechanism requires sufficiently high magnetization near the horizon, and it is therefore anticipated that overloading of magnetic field lines by relativistic pairs may lead to a complete shutdown of this process.    In the latter case, the outflow will be driven by the pressure of the injected plasma, as described in Section \ref{sec:model}, rather than by magnetic extraction of the spin energy of the black hole.  Interestingly, relatively low accretion rates may be favorable for producing powerful outflows.
The critical load above which activation of the BZ process is prevented is a key issue, currently under investigation (Globus \& Levinson, in preparation).

\newpage

\end{document}